# Few-Shot Speaker Identification Using Lightweight Prototypical Network with Feature Grouping and Interaction

Yanxiong Li, *Senior Member, IEEE*, Hao Chen, Wenchang Cao, Qisheng Huang and Qianhua He, *Senior Member, IEEE*

*Abstract*—Existing methods for few-shot speaker identification (FSSI) obtain high accuracy, but their computational complexities and model sizes need to be reduced for lightweight applications. In this work, we propose a FSSI method using a lightweight prototypical network with the final goal to implement the FSSI on intelligent terminals with limited resources, such as smart watches and smart speakers. In the proposed prototypical network, an embedding module is designed to perform feature grouping for reducing the memory requirement and computational complexity, and feature interaction for enhancing the representational ability of the learned speaker embedding. In the proposed embedding module, audio feature of each speech sample is split into several low-dimensional feature subsets that are transformed by a recurrent convolutional block in parallel. Then, the operations of averaging, addition, concatenation, element-wise summation and statistics pooling are sequentially executed to learn a speaker embedding for each speech sample. The recurrent convolutional block consists of a block of bidirectional long short-term memory, and a block of de-redundancy convolution in which feature grouping and interaction are conducted too. Our method is compared to baseline methods on three datasets that are selected from three public speech corpora (VoxCeleb1, VoxCeleb2, and LibriSpeech). The results show that our method obtains higher accuracy under several conditions, and has advantages over all baseline methods in computational complexity and model size.

*Index Terms*—Few-shot learning, speaker identification, feature grouping, feature interaction, prototypical network

## I. Introduction

SPEAKER recognition is a quite critical technique for many practical applications, such as criminal investigation [1], financial services [2]. It can be mainly divided into two classes: speaker identification (SI) and speaker verification (SV) [3]. The SI is a task to decide which speaker utters a given speech sample [4], while the SV is a task to reject or accept the identity claim of a speaker based on the speaker's voice [5].

In some applications, it is difficult to acquire enough speech samples for building a reliable SI system. For instance, law enforcement agencies often have difficulty in collecting enough speech samples spoken by the suspects for forensic SI [1]. However, few samples are needed for humans to learn a new task well [6]. To reduce the performance gap between machine learning and human learning when the speech samples are very few, a task of FSSI is proposed [7], [8]. The FSSI is a newly emerging task to identify speakers in unlabeled speech samples (query set) based on few labeled speech samples (support set).

In this paper, we propose a FSSI method using a lightweight prototypical network with feature grouping and interaction. The rest of this paper is structured as follows. Sections II and III describe related works and our contributions, respectively. Section IV introduces the proposed FSSI method. Section V presents the experiments and discussions, and the conclusions are drawn in Section VI.

## II. Related Works

In this section, we introduce related works from three aspects, including general speaker recognition, few-shot speaker recognition, and lightweight speaker recognition.

### A. General Speaker Recognition

Many efforts were made on speaker recognition. They mainly focused on solving two problems: how to learn a front-end feature with strong representational ability, and how to build a back-end classifier with high accuracy for recognition.

Hand-crafted features were designed to represent properties of different speakers, such as constant Q cepstral coefficients [9], Mel-frequency cepstral coefficients [10], linear prediction coding coefficients [10], Gaussian supervector [11], eigenvoice motivated vectors [12], and I-vector [13]. Each one of these features was often designed for a specific scenario. Hence, they would not perform well in other scenarios. In addition, they are shallow-model based features, instead of deep-model based features. They cannot effectively represent the differences of deep-level properties among various speakers. To overcome the shortcomings of hand-crafted features, deep neural network (DNN) was proposed to learn deep-model based features. The DNN can learn discriminative embeddings from speech samples. Hence, the deep-model based features exceeded the hand-crafted features for speaker recognition. The deep-model based features include the X-vector learned by a time-delay neural network (TDNN) [14], [15]; the d-vector learned by a DNN [16]; and the S-vector learned by a Transformer encoder speaker authenticator [17]. In addition, other networks were used to learn or adapt embeddings for speaker recognition, such as long short-term memory network [18], convolutional neural network [19], Siamese neural network (SNN) [20], Transformer [21], and multi-scale convolutional recurrent network [22].

On the other side, many works were also done on the design of back-end classifiers. Typical classifiers adopted in previous works mainly include: DNN [23], vector quantization [24], Gaussian mixture model [25], support vector machine [26], hidden Markov model [27], probabilistic linear discriminant analysis (PLDA) [28], and cosine distance [10].

This work was supported by national natural science foundation of China (62111530145, 61771200), international scientific research collaboration project of Guangdong (2021A0505030003), Guangdong basic and applied basic research foundation (2021A1515011454, 2022A1515011687).

All authors of this paper are with the School of Electronic and Information Engineering, South China University of Technology, Guangzhou, China. The corresponding author is Dr. Yanxiong Li (eeyxli@scut.edu.cn).

## B. Few-Shot Speaker Recognition

Since the DNN requires a huge amount of training data for achieving satisfactory results, a few works were done for few-shot speaker recognition with various DNNs. In these works, a common practice is to obtain an end-to-end neural network for speaker embedding learning [7], [29], [30]. For instance, Wang et al. [7] built an end-to-end neural network to learn speaker embedding by a prototypical loss for few-shot speaker recognition. Recently, Wang et al. [31] proposed an end-to-end neural network with an attention corrected prototype using the relation based indefinite distance metric for few-shot speaker recognition. In addition, Li et al. [32] designed a depthwise separable convolutional network with channel attention for FSSI. Their neural network was trained with a prototypical loss, which can alleviate the overfitting problem.

The technique of adversarial learning [33] is to generate two neural networks in an adversarial way for enhancing training efficiency, which is beneficial for few-shot speaker recognition. For example, Li et al. [34] designed an adversarial model for few-shot speaker recognition. The technique of transfer learning [35] was used for few-shot speaker recognition too. For instance, Anand et al. [8] utilized a capsule network [36] for few-shot speaker recognition. In addition, the technique of meta-learning was applied for few-shot speaker recognition. For example, Li et al. [37] built the bridging mixture density networks with meta learning to identify speakers using a few samples. Kye et al. [38] adopted the meta-learning to tackle the problem of imbalance length of training and testing samples. Mishra et al. [39] used a SNN [40] and 3-dimensional convolution to tackle the problem of few-shot speaker verification.

These methods above for few-shot speaker recognition can solve the problem of performance degradation caused by the lack of samples, but model size and computational complexity of these methods are relatively high and are not explicitly considered in previous works. It is very challenging to deploy these methods on intelligent terminals with limited resources, such as smart watches, smart speakers, service robots.

## C. Lightweight Speaker Recognition

To deploy speaker recognition systems on intelligent terminals with limited resources, it is necessary to study the problems of either lightweight model design or model compression while maintaining the performance of speaker recognition. These problems belong to the scope of lightweight speaker recognition.

Inspired by the successes of MobileNet [41] and ShuffleNet [42] in computer vision, some efforts were made to improve convolutional operations for obtaining a lightweight model for speaker recognition [43], [44]. For example, Koluguri et al. [43] proposed a SpeakerNet for the tasks of speaker recognition and verification. The SpeakerNet is composed of residual blocks with 1-dimensional depth-wise separable convolutions, batch normalization, and ReLU layers, which is a lighter model with 5 million parameters. Similarly, Nunes et al. [44] proposed a portable model called additive margin MobileNet1D for implementing speaker identification on mobile devices. The MobileNet1D takes 11.6 megabytes on disk storage. In short, such improved convolutional operations can reduce the model size and computational complexity, but they generally lead to different levels of performance degradation. In addition, the tradeoff between the complexity reduction and performance degradation requires to be carefully considered.

Some works were made on manual design of a model with better architecture or organization for realizing lightweight speaker recognition [45]-[48]. For example, inspired by the success of residual network (ResNet) in image recognition [49], Oneață et al. [45] replaced the original trunk of the SincNet [50] with a lightweight (ResNet-inspired) trunk. They found that their trunk was lighter (2.8 million instead of 22 million parameters) with better performance. Recently, inspired by the favorable geometry of the hyperbolic geometry, Lee et al. [48] proposed a lighter model called hyperbolic ResNet for speaker recognition. They found that the learned speaker embeddings were more compact and were at the same level of performance. In summary, the minimum size of all such proposals for designing a model with better architecture or organization is at the level of millions of parameters. Therefore, there is still much room to reduce the size of the model in these methods.

In addition, knowledge distillation [51] and neural architecture search (NAS) [52] were used to realize lightweight speaker recognition [53]-[55]. For instance, Ng et al. [53] investigated the framework of teacher-student training for knowledge distillation in the text-independent speaker recognition, and obtained competitive result with 88-93% smaller models. Lin et al. [54] proposed an asymmetric structure, which took a big model of the ECAPA-TDNN (Emphasized Channel Attention, Propagation and Aggregation in TDNN) for enrollment and a small-scale model of the ECAPA-TDNN-Lite for verification. The ECAPA-TDNN-Lite obtained competitive equal error rates with 11.6 million FLOPS (floating-point operations per second). Wang et al. [55] used the NAS to automatically design an efficient network (EfficientTDNN) which consisted of a TDNN based super-net and a TDNN-NAS algorithm. Their neural network obtained competitive equal error rates with Multiply-Accumulate operations (MACs) from 204 million to 1.45 giga. Although these methods can obtain a relatively small model for testing, they require to pre-train a very large model or to search the proper network with large number of architecture settings. Hence, the training cost and overall requirement for realizing speaker recognition are still heavy.

## III. OUR CONTRIBUTIONS

Based on the descriptions above, it can be known that many efforts are made on few-shot or lightweight speaker recognition. However, to our best knowledge, there is almost no work on lightweight few-shot speaker recognition until now.

The work in this paper focuses on the SI task only. We propose a FSSI method by a lightweight prototypical network in which an embedding module is designed to realize feature grouping and interaction. In the embedding module, the input feature with $H$-dimension is evenly split into $I$ feature subsets with $J$-dimension. Afterwards, these $I$ feature subsets are independently fed into a recurrent convolutional block (RCB) which consists of a block of bidirectional long short-term memory (BLSTM) and a block of de-redundancy convolution (DRC). Then, some typical operations (e.g., averaging, addition, concatenation, statistics pooling) are sequentially executed on the output of the RCB to produce a speaker embedding for each speech sample. Finally, a Softmax layer is introduced to the proposed prototypical network for realizing the FSSI task.

Three datasets are constructed by randomly selecting speech samples from three speech corpora (VoxCeleb1, VoxCeleb2, and LibriSpeech) for evaluating various methods. The results show that our method is effective for lightweight FSSI. In conclusion, the main contributions of this work are as follows:

1. To efficiently learn the speaker embeddings with strong representational ability, we design an embedding module mainly consisting of a RCB to implement feature grouping and interaction. The RCB is composed of a BLSTM block and a DRC block, which can capture global sequential information and local spatial information during speaker embedding learning. To the best of our knowledge, the architecture of the proposed embedding module (including the RCB and DRC) is novel and not used in previous works.
2. The idea of both feature grouping and feature interaction for speaker embedding learning is not considered and used in previous works for speaker recognition. The operation of feature grouping can reduce the model size and computational complexity, while the operation of feature interaction can enhance the representational ability of the learned speaker embedding. In addition, these two operations are executed twice during the speaker embedding learning. The first time is conducted at the preceding and succeeding blocks of the RCB, while the second time is performed at the DRC block. In short, the adoption of these two operations is the most critical reason why our proposed embedding module is effective for lightweight FSSI and achieves better performance.
3. We propose a FSSI method using a lightweight prototypical network for solving the problem of lightweight FSSI which is not considered in previous works. We thoroughly validate the effectiveness of our proposed method, and compare it with the baseline methods on three experimental datasets under different conditions. Experimental results show that our proposed method has advantages over the baseline methods in accuracy, model size and computational complexity.

## IV. METHOD

In this section, we describe the proposed method, including the descriptions of problem definition, network architecture, embedding module, and recurrent convolutional block.

### A. Problem Definition

In this study, we focus on investigating the problem of text-independent FSSI by an episode-based strategy [56]. In each episodic training step, $K$ samples from each one of $N$ classes (speakers) are randomly selected from training subset as the support set (i.e., $K \cdot N$ support samples), and then another $K$ samples from each one of the $N$ speakers are also randomly selected from the remaining samples of the training subset to form the query set (i.e., $K \cdot N$ query samples). The $N$ speakers in the support set are the same as those in the query set, whereas the $K \cdot N$ support samples are completely different from the $K \cdot N$ query samples. These $2K \cdot N$ samples form one batch of support and query samples that are fed into the prototypical network in one episodic learning procedure. In each episodic testing step, the speaker of each query sample is identified after a set of support samples are fed into the network for enrollment. It is supposed that each query sample belongs to one of the speakers in the support set during the episodic testing step. We evaluate on the variable numbers of both support speakers ($N$-way) and support samples ($K$-shot). This is the $N$-way $K$-shot task.

After extracting speaker embeddings from support and query samples by the embedding module of the network, the support set and query set are denoted as $S = \{X_1, …, X_n, …, X_N\}$ and $Q = \{X_1, …, X_n, …, X_N\}$, respectively. Furthermore, $X_n = \{(x_{n,1}, y_n), …, (x_{n,k}, y_n), …, (x_{n,K}, y_n)\}$ denotes the support (or query) set of the $n$th speaker, where $1 \leq n \leq N$, $1 \leq k \leq K$. In each $X_n$, $x_{n,k}$ and $y_n$ denote the $k$th speaker embedding and the speaker label of the $n$th speaker in support (or query) set, respectively.

The prototype vector $\bar{x}_n$ of the $n$th speaker is defined by the mean vector of speaker embeddings in the support set of the $n$th speaker, namely,

$$\bar{x}_n = \frac{1}{K}\sum_{k=1}^{K} x_{n,k} . \qquad (1)$$

The probability produced by the prototypical network over speakers for a query speaker embedding $x$ based on a Softmax over distances to the prototype vectors is defined by

$$p(y = y_n \mid x) = \frac{\exp(-Dist(\bar{x}_n, x))}{\sum_{n'} \exp(-Dist(\bar{x}_{n'}, x))}, \qquad (2)$$

where $\bar{x}_n$ and $\bar{x}_{n'}$ denote prototype vectors of the $n$th speaker in support set; $x$ and $y$ denote one speaker embedding in query set and the predicted speaker label, respectively; $Dist(\cdot)$ denotes a distance function, such as Euclidean distance, cosine distance.

Learning proceeds by minimizing a loss function $\ell$ via the stochastic gradient descent algorithm [57]. The loss function $\ell$ is defined by

$$\ell = -\log(p(y = y_n \mid x)). \qquad (3)$$

A prototypical network is trained to minimize the loss function $\ell$ by repeatedly feeding batches of speaker embeddings of both support and query samples to the network, so that each query sample corresponds to one of the speakers in the support set.

### B. Network Architecture

The proposed prototypical network for FSSI is illustrated in Fig. 1, which is mainly composed of two parts: one embedding module and one Softmax layer. The embedding module is designed to learn speaker embedding from support samples and query samples, while the Softmax layer is adopted to make a decision for speaker identification.

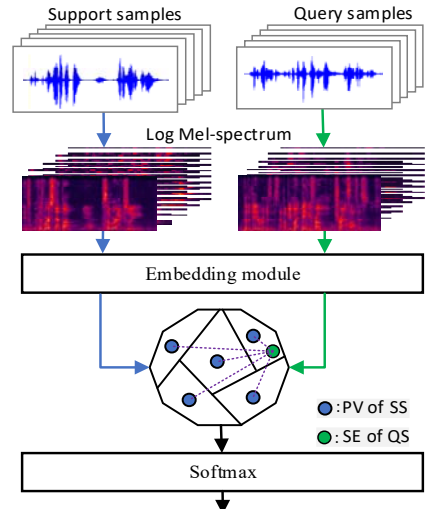

Fig. 1. The proposed prototypical network for FSSI. PV: prototype vectors; SS: support samples; SE: speaker embeddings; QS: query samples.

The motivation for designing the embedding module is based on two reasons. First, key part of the embedding module is the RCB which is a recurrent convolutional architecture (BLSTM + DRC). The embedding module can capture global sequential information by the BLSTM block from the log Mel-spectrum [58] and local spatial information by the DRC block from the transformed feature (output by the BLSTM block). These two kinds of information are complementary to each other, and will enhance the representational ability of speaker embedding. In addition, the embedding learned by recurrent convolutional neural network exceeds the embeddings that are learned by recurrent neural network or convolutional neural network for other audio processing tasks [59], [60]. Second, the embedding module performs feature grouping and feature interaction during speaker embedding learning. The feature grouping can decrease the model size and computational complexity. Meanwhile the feature interaction can capture correlation information among feature subsets and is helpful for improving the representational ability of speaker embedding.

The motivation for using the Softmax layer is based on two reasons. First, the Softmax layer can be stacked at the top of the embedding module to seamlessly form a prototypical network with discriminability for realizing the FSSI task. Second, the network with the proposed embedding module can be efficiently updated under the guidance of the loss function $\ell$ without tuning complex hyper-parameters.

### C. Embedding Module

The framework of the embedding module is depicted in Fig. 2.

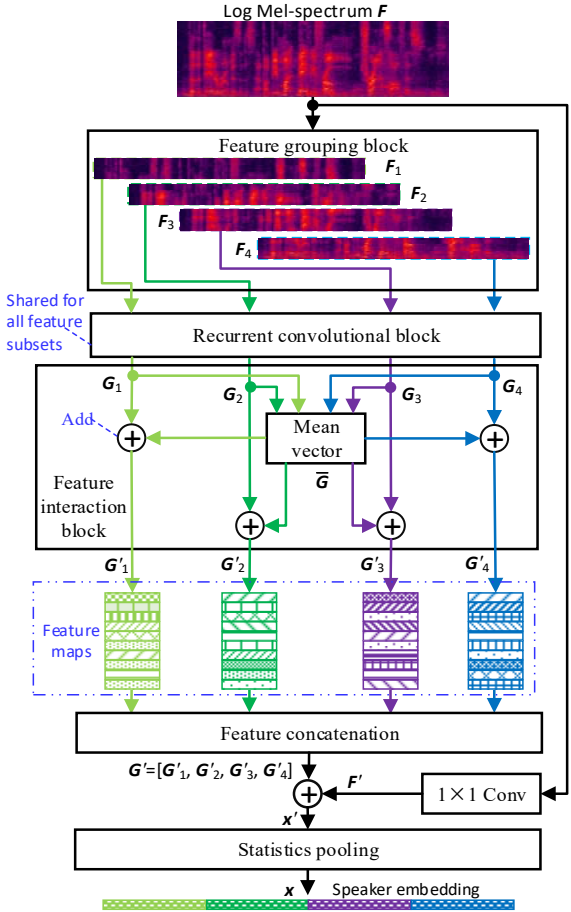

Fig. 2. The framework of embedding module for learning speaker embedding.

The input audio feature of log Mel-spectrum $F \in \mathbb{R}^H$ [58], is extracted from each support sample or query sample. Along the frequency-dimension, the log Mel-spectrum is split into $I$ (taking 4 as an example as shown in Fig. 2) feature subsets $F_i \in \mathbb{R}^J$, where $H=IJ$, $I \in \mathbb{Z}^+$, $J \in \mathbb{Z}^+$ and $1 \leq i \leq I$. $I$ feature subsets $F_i$, instead of $F$, are independently fed into the RCB for further transformation. That is, the RCB is shared for $I$ feature subsets $F_i$. Compared with the feature $F$, the feature subset $F_i$ is with lower dimension. Hence, the number of parameters of the RCB (e.g., numbers of hidden units of BLSTM and convolutional kernels) can be reduced when the $F_i$ is used as the input of the RCB. However, the correlation information among $I$ feature subsets cannot be utilized when each $F_i$ is independently fed into the RCB. As a result, the representational ability of the learned speaker embedding will be weakened.

To acquire the correlation information among $I$ feature maps (subsets) $G_i$, we design a block of feature interaction which consists of two operations: the calculation of mean vector $\bar{G}$, and the addition of both mean vector $\bar{G}$ and feature maps $G_i$. $I$ transformed feature maps $G_i=\{g^i_{m,n}\}$ are fed into the block of "Mean vector" for calculating the mean vector $\bar{G}=\{\bar{g}_{m,n}\}$ by

$$\bar{g}_{m,n} = \frac{1}{I} \sum_{i=1}^{I} g^i_{m,n}, \qquad (4)$$

where $1 \leq m \leq h$, $1 \leq n \leq w$; $h$ and $w$ denote the height and width of the feature maps $G_i$, respectively. The mean vector $\bar{G}$ is a transformation of all feature maps $G_i$, and thus contains the correlation information among all $G_i$. Each $G_i$ is further transformed to $G'_i = \{g'^i_{m,n}\}$ by adding both $G_i$ and $\bar{G}$, namely

$$g'^i_{m,n} = g^i_{m,n} + \bar{g}_{m,n}. \qquad (5)$$

$I$ feature maps $G'_i$ are concatenated together to obtain a transformed feature map $G'=[G'_1, \ldots, G'_i, \ldots, G'_I]$. The log Mel-spectrum $F$ is convoluted with convolution kernels with the size of $1 \times 1$ (i.e., $1 \times 1$ Conv in Fig. 2) to obtain transformed feature map $F'$. An operation of statistics pooling [15] is conducted on the addition of both $G'$ and $F'$, namely $x' = G' + F'$, to produce speaker embedding $x$. Specifically, in the statistics pooling layer, both mean vector $x'_{mean}$ and standard deviation vector $x'_{std}$ of $x'$ are calculated and then the $x'_{mean}$ and $x'_{std}$ are concatenated to form the speaker embedding $x=[x'_{mean}, x'_{std}]$.

### D. Recurrent Convolutional Block

The RCB is composed of one BLSTM block and one DRC block, as shown Fig. 3.

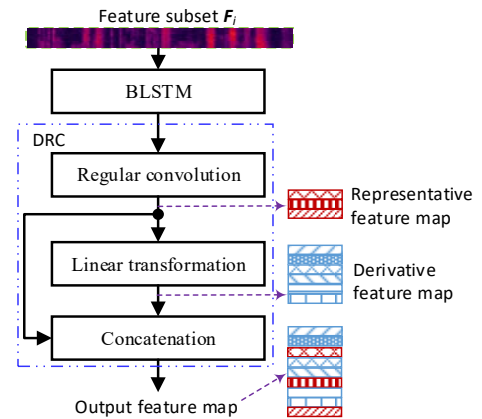

Fig. 3. The recurrent convolutional block.

The BLSTM block consists of one long short-term memory block (LSTMB) and one bidirectional recurrent block (BRB), which can acquire the global sequential information from two directions with forward layer and backward layer [61], [62]. The global sequential information needs to be captured for speaker identification with better performance. In addition, deep convolutional neural network (DCNN) consists of many regular convolutions that leads to heavy computational load with occupying large memory [63]. There is much redundant information in the feature maps output by convolutional layers of the DCNN, and most of these feature maps are similar to each other. The output feature map $Z$ at a regular convolutional layer is defined as the convolution of the input feature map $Y$ and the convolution kernel $K$, namely

$$Z = Y * K, \qquad (6)$$

where $Z \in \mathbb{R}^{h' \times w' \times M}$, $Y \in \mathbb{R}^{C \times h \times w}$, $K \in \mathbb{R}^{C \times h' \times w' \times M}$; $*$ denotes the convolutional operation; $h'$ and $w'$ denote the height and width of the output feature map, respectively; $h$ and $w$ stand for the height and width of the input feature map, respectively; $M$ and $C$ represent the numbers of output channels and input channels, respectively. In Eq. 6, the bias vector is omitted for simplicity.

In practice, it is dispensable to produce all feature maps using regular convolutions. It will be computationally efficient and of low redundancy, if some representative feature maps are produced by regular convolutions and the remaining (derivative) feature maps are generated using some linear transformations with efficient computation based on the representative feature maps. It is supposed that $L$ representative feature maps $Z'$ are produced by regular convolutions and defined by

$$Z' = Y * K', \qquad (7)$$

where $Z' \in \mathbb{R}^{h' \times w' \times L}$, $K' \in \mathbb{R}^{C \times h' \times w' \times L}$; $L$ is the number of output channels and $L \leq M$; and the bias vector is omitted for simplicity.

To generate the rest ($M$-$L$) output feature maps, linear transformations are conducted on each representative feature map to produce its derivative feature maps. These linear transformations are defined by

$$Z^d_{l,e} = \psi_{l,e}(Z'_l), \qquad (8)$$

where $l \in \mathbb{Z}^+$, $1 \leq l \leq L$; $e \in \mathbb{Z}^+$, $1 \leq e \leq E-1$, $E = M/L$; $\psi_{l,e}(\cdot)$ is the $e$th linear transformation which is conducted on the $l$th representative feature map $Z'_l$ to produce the $e$th derivative feature map $Z^d_{l,e}$.

Afterwards, the $l$th representative feature map and its $E$-1 derivative feature maps $Z^d_{l,e}$ are concatenated for generating $E$ output feature maps $Z^o_{l,\cdot} = [Z^d_{l,1}, Z^d_{l,2}, \ldots, Z^d_{l,E-1}, Z'_l]$. That is, $Z^o_{l,e} = Z^d_{l,e}$, $1 \leq e \leq E-1$; and $Z^o_{l,E} = Z'_l$. The linear transformation above is conducted on all representative feature maps and then $M = LE$ output feature maps $Z^o = [Z^o_{1,\cdot}, Z^o_{2,\cdot}, \ldots, Z^o_{L,\cdot}]$ are obtained. It should be noted that computational load of the linear transformations defined by Eq. 8 is much lighter than that of the regular convolutions defined by Eq. 7. The linear transformations can be implemented in different ways, such as depthwise convolution, linear weighting, wavelet transformation. Depthwise convolution is used here for realizing the linear transformations, since it is supported well by software (toolkits) and hardware and thus is computationally efficient [41].

In short, the DRC block can acquire local spatial information with low redundancy by both regular convolutions and linear transformations. The $M$ output feature maps consist of two groups: $L$ representative feature maps and $L \cdot (E-1)$ derivative feature maps. That is, there is feature grouping in the DRC block. In addition, the derivative feature maps are produced by performing linear transformations on the representative feature maps. Namely, there is feature interaction in the DRC block. Hence, the acquisition of the output feature maps includes the operations of feature grouping and feature interaction as well.

## V. EXPERIMENTS AND DISCUSSIONS

In this section, we first describe experimental data and setups. Afterwards, we discuss ablation experiments and conduct a qualitative analysis for our method. Then, we compare our method with the state-of-the-art methods and discuss generalization of different methods across various datasets. Finally, we compare different methods for speaker verification.

### A. Experimental datasets

Experiments are conducted on the datasets that are selected from three public speech corpora, including VoxCeleb2 [64], VoxCeleb1 [65], and Librispeech [66]. In addition, they are three of the most common large-scale speech corpora used in previous works for speaker recognition.

The VoxCeleb2 has 1128246 speech samples spoken by 6112 speakers. Its total duration is 2442 hours. It is roughly gender balanced with 61% of male speakers. The speakers span a wide range of various ethnicities, accents, professions, and ages. Its speech samples are degraded with background chatter, laughter, overlapping speech, and different room acoustics.

The VoxCeleb1 consists of more than 100 thousand speech samples that are uttered by 1251 speakers. Total length of the VoxCeleb1 is 352 hours. The speech samples are collected from videos on the YouTube website. There are many different speakers in this corpus, spanning different races, accents, occupations and ages. The classes of samples in this corpus include indoor studios and outdoor stadiums, speeches given to large audiences, interviews from red carpets, excerpts from shot multimedia, and so on. The real-world noise contained in the samples includes background utterances (e.g., chatter, laughter), noise of recording equipment, room reverberations.

The LibriSpeech is a collection of approximately 1000 hours of audiobooks that are uttered by 2484 speakers. Training data is split into 3 parts of 100 hours, 360 hours, and 500 hours. Development and testing data are split into the *clean* and *other* categories, respectively. Each of the development and testing subsets is about 5 hours in length.

Experimental datasets that are extracted from the VoxCeleb2, the VoxCeleb1 and the LibriSpeech, are denoted as V2-set, V1-set, and L-set, respectively. They are independently divided into training and testing subsets without overlaps. The length of each sample is 7 seconds. Their information is listed in Table I.

TABLE I
DETAILED INFORMATION OF EXPERIMENTAL DATASETS

| Parameters | V2-set | | V1-set | | L-set | |
| --- | --- | --- | --- | --- | --- | --- |
|  | Training | Testing | Training | Testing | Training | Testing |
| # Speakers | 2000 | 300 | 951 | 300 | 921 | 300 |
| # Samples | 40000 | 6000 | 19020 | 6000 | 18420 | 6000 |
| Length (hours) | 77.78 | 11.67 | 36.98 | 11.67 | 35.82 | 11.67 |
| #Samples/Spkr | 20 | 20 | 20 | 20 | 20 | 20 |

# Samples/Spkr: number of samples per speaker.

In each episodic training step, $N \cdot K$ samples are randomly selected from $N$ speakers ($K$ samples per speaker) in the training subset to generate the support set, and then another $N{\cdot}K$ different samples of the same $N$ speakers ($K$ samples per speaker) are randomly chosen from the training subset to form the query set. In each episodic testing step, the selections of query samples and support samples from the testing subset are the same as that in the episodic training step. The selection procedures of speakers and speech samples per speaker are repeated until all speakers and their speech samples in the training and testing subsets are selected. The selected speech samples in various episodes are different to each other. The average score of the repeated results in episodic testing steps is used as the final result of this test. In addition, we construct the testing subset ten times by randomly selecting speech samples from each speech corpus, and conduct ten tests. The final results are the average of the ten test results.

*B. Experimental Setup*

Our experiments are done on a machine whose configurations are as follows: a CPU of Intel(R) Core (TM) i7-6700 with 3.10 GHz, a RAM of 64 GB, and a GPU of NVIDIA 1080 TI. All experiments are performed by the toolkit of PyTorch.

The metric of accuracy is used to measure the identification performances of various methods, which is defined as the ratio of the number of correctly identified samples to the total number of samples. The higher the value of accuracy, the better the identification performance of the methods. In addition, memory requirement and computational complexity of various methods are measured by the metrics of model size (MS) and MACs, respectively. The MS is defined as the total number of parameters of a neural network. The MACs is defined as the number of multiplication and addition operations of a neural network. The lower the value of MACs (or MS), the lower computational complexity (or memory requirement) of the methods. Main parameters of our method are listed in Table II.

TABLE II
MAIN PARAMETER SETTINGS OF OUR METHOD

| Type | Parameters settings |
|---|---|
| Preprocessing | Frame length/overlapping: 25ms/10ms<br>Dimension of log Mel-spectrum: 80 |
| Prototypical network | No. of feature subsets, $I$: 1 to 16<br>No. of hidden units of BLSTM: 40<br>No. of convolutional kernels: 32<br>Size of convolutional kernels: 3×3<br>No. of output feature maps of DRC, $M$: 256<br>No. of representative feature maps of DRC, $L$: 128<br>Dimension of speaker embedding: 512<br>Distance function in Eq. 2: Euclidean distance |

*C. Ablation Experiments*

In this subsection, we discuss the settings of two parameters that have direct influences on the performance of our method. These two parameters are the number of the feature subsets (i.e., $I$), and the ratio of the number of output feature maps to the number of representative feature maps (i.e., $M/L$). In the experiments, the value of $N$-way $K$-shot is set to 5-way 5-shot.

We discuss the performance of our method with different numbers of feature subsets. The number of feature subsets, $I$, ranges from 1 to 16 in steps of power of 2. On the other hand, the ratio of $M/L$ is set to 2 and other parameters are configured as given in Table II. The scores of MS, MACs and accuracy obtained by our method with different numbers of feature subsets are presented in Table III.

TABLE III
RESULTS OF OUR METHOD WITH DIFFERENT NUMBERS OF FEATURE SUBSETS

| $I$ | MS (Kilo) | MACs (Million) | V2-set Acc. (%) | V1-set Acc. (%) | L-set Acc. (%) |
|---|---|---|---|---|---|
| 1 | 146.30 | 103.16 | 92.65 | 92.30 | 98.29 |
| 2 | 84.86 | 60.21 | 92.69 | 92.25 | 97.43 |
| 4 | 54.14 | 38.74 | **92.89** | **92.74** | **98.51** |
| 8 | 38.78 | 28.00 | 92.05 | 92.47 | 98.27 |
| 16 | 31.10 | 22.64 | 91.75 | 92.38 | 97.94 |

Acc.: Accuracy.

When the number of feature subsets is equal to 4, our method obtains the highest accuracy scores of 92.89%, 92.74%, and 98.51% on the V2-set, V1-set, and L-set, respectively, with relatively low values of both MS and MACs. When the number of feature subsets deviates from 4, the accuracy scores steadily decrease. Hence, the number of feature subsets is set to 4 in the experiments of the following sections.

With the increase of the number of feature subsets (e.g., from 1 to 4), the dimension of each feature subset which is fed into the RCB becomes lower. As a result, the number of parameters required for the RCB is reduced, and the neural network becomes lightweight. Thanks to the operation of feature interaction executed in the embedding module, the correlation information among all feature subsets is effectively captured for enhancing the representational ability of the learned speaker embedding. Hence, the accuracy score obtained by our method is increased. However, when the input feature is segmented into too many feature subsets (e.g., >4), the correlation information among all feature subsets is too fragmented to be effectively acquired by the operation of feature interaction. Hence, the representational ability of the learned speaker embedding is weakened. As a result, the accuracy score obtained by our method is reduced and lower than that when $I$ is equal to 4.

In addition, we discuss the impact of the ratio of $M/L$ on the performance of our method. The values of the ratio of $M/L$ range from 1 to 4. In this experiment, the number of feature subsets is set to 4 and other parameters are configured as shown in Table II. The scores of MS, MACs and accuracy obtained by our method with different values of $M/L$ are listed in Table IV.

TABLE IV
RESULTS OF OUR METHOD WITH VARIOUS RATIOS OF THE NUMBER OF OUTPUT FEATURE MAPS TO THE NUMBER OF REPRESENTATIVE FEATURE MAPS ($M/L$)

| $M/L$ | MS (Kilo) | MACs (Million) | V2-set Acc. (%) | V1-set Acc. (%) | L-set Acc. (%) |
|---|---|---|---|---|---|
| 1 | 79.12 | 56.23 | 92.44 | 92.38 | 98.11 |
| 2 | 54.14 | 38.74 | **92.89** | **92.74** | **98.51** |
| 3 | 53.31 | 38.25 | 92.71 | 92.65 | 98.39 |
| 4 | 51.85 | 37.12 | 92.41 | 92.51 | 98.18 |

When the value of $M/L$ is equal to 1, the output feature maps of the DRC consist of representative feature maps only (without derivative feature maps). That is, there is regular convolution only (without linear transformation) and thus without the interactions of representative feature maps in the DRC. In this case, our method obtains satisfactory results on three datasets in accuracy, but the values of MS and MACs reach the maximum.

When the value of $M/L$ is equal to 2, our method obtains the highest accuracy scores of 92.89%, 92.74%, and 98.51% on the V2-set, V1-set, and L-set, respectively, with relatively low values of both MS and MACs. With the increase of the value of $M/L$, the proportion of the derived feature maps in the output feature maps increases and thus the neural network becomes lighter, but the accuracy scores decrease. Hence, the value of $M/L$ is set to 2 in the following experiments.

*D. Qualitative Analysis*

In this subsection, we make a qualitative analysis about the influence of feature interaction on the representational ability of the learned feature subsets. The t-SNE [67] is utilized to map the feature subsets $G_i$ and $G'_i$ into a two-dimensional space. $G_i$ and $G'_i$ ($1 \leq i \leq I$) are the input and output feature subsets of the feature interaction block of the proposed embedding module, respectively, as given in Fig. 2 (where $I=4$). We adopt the Python library of *scikit-learn* to reduce the dimensionality of $G_i$ and $G'_i$, and utilize the Python library of *matplotlib* to plot the distributions of $G_i$ and $G'_i$ in the two-dimension space. Without loss of generality, five speakers (5-way) are randomly selected from the V2-set for demonstrating the distributions of their corresponding feature subsets. The distributions of $G_i$ and $G'_i$ in the two-dimensional space are depicted in Fig. 4 where $1 \leq i \leq 2$ for simplicity.

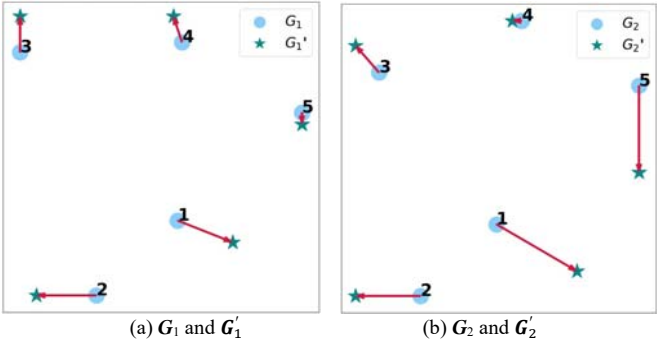

Fig. 4. Visualization of the feature subsets $G_i$ and $G'_i$ ($1 \leq i \leq 2$). $G_i$ and $G'_i$ are the input and the output of the feature interaction block of the embedding module, respectively. The digits from 1 to 5 represent five different speakers.

It can be seen from Fig. 4 (a) and Fig. 4 (b) that the distance between the feature subsets $G'_i$ of different speakers is greater than that between the feature subsets $G_i$ of different speakers. That is, compared to the feature subsets $G_i$ (without feature interaction), the feature subsets $G'_i$ (with feature interaction) are shifted away from the confusion region in the two-dimensional space to obtain more discriminative decision boundaries. Accordingly, the confusion between the five $G'_i$ are expected to be less than that between the five $G_i$. In other words, after transformation by the feature interaction block of the proposed embedding module, the representation ability of the feature subsets can be improved.

*E. Comparison of Different Methods in Accuracy*

In this subsection, our method is compared to ten baseline FSSI methods by the episode-based strategy. These baseline methods are briefly described as follows.

The method in [44] was inspired by the success of the MoiblieNet [41], where a portable model of MobileNet1D (MN1D) was used for speaker recognition. The technique of group convolution (GC) [63] was proposed to reduce the complexity of neural networks. In the GC based method for speaker identification, each input feature map was split into some sub-feature-maps to execute the operation of convolution in each convolutional layer. Afterwards, the sub-feature-maps after convolution were concatenated to obtain the output feature maps for further transformations. Lin et al [54] designed an ECAPA-TDNN-Lite (ETL) model using the knowledge distillation for lightweight speaker recognition. A prototypical network (PN) was proposed by Snell et al. [56] for few-shot classification. The method in [68] was a model agnostic meta learning (MAML) algorithm and was used for FSSI. Vinyals et al. [69] proposed a matching network (MN) for few-shot learning. Snyder et al. [14] proposed a framework of the TDNN to learn the X-vector for speaker recognition. Gao et al. [70] designed a multi-scale convolutional neural network, namely Res2Net, where multi-scale convolution instead of single-scale convolution was adopted to learn speaker embedding. The technique of principal component analysis (PCA) was discussed in [71] for reducing the dimensionality of the input feature. In addition, a filter-based feature selection method (filter based) [72] was employed for selecting dominant feature subsets from the original feature. In the Res2Net based, PCA based and filter based methods, the input feature was processed by the modules of multi-scale convolution, dimensionality reduction and feature selection, respectively. Afterwards, the processed feature subsets in these three methods were fed into a back-end prototypical network for making decision. Based on the descriptions above, main technical merits of different methods are listed in Table V.

TABLE V
SUMMARY OF DIFFERENT METHODS FOR FEW-SHOT SPEAKER IDENTIFICATION

| Methods | Main merits |
|---|---|
| Ours | Feature grouping and interaction |
| MN1D based [44] | Depthwise separable convolution |
| GC based [63] | Group convolution |
| ETL based [54] | Knowledge distillation |
| PN based [56] | Prototypical loss |
| MAML based [68] | Meta learning |
| MN based [69] | Metric learning |
| TDNN based [14] | Frame-level and segment-level modeling |
| Res2Net based [70] | Multi-scale convolution |
| PCA based [71] | Principal component analysis |
| Filter based [72] | Feature selection |

Main parameters of all baseline methods are set according to the recommendations in the respective references and are optimally tuned on training data. We compare the performance of different methods on three datasets when the values of $N$-way $K$-shot are set to 5-way 5-shot, 5-way 1-shot, 10-way 5-shot, and 10-way 1-shot. Under the same conditions, the accuracy scores that are obtained by different methods on the datasets of the V2-set, V1-set and L-set are presented in Table VI, Table VII and Table VIII, respectively.

TABLE VI
ACCURACY SCORES OBTAINED BY DIFFERENT METHODS ON THE V2-SET (IN %)

| Methods | 5-way 5-shot | 5-way 1-shot | 10-way 5-shot | 10-way 1-shot |
|---|---|---|---|---|
| Ours | **92.89** | **78.31** | **88.87** | **68.88** |
| MN1D based | 92.27 | 75.32 | 88.18 | 66.92 |
| GC based | 92.33 | 76.02 | 88.20 | 67.00 |
| ETL based | 90.83 | 74.04 | 85.70 | 64.25 |
| PN based | 90.47 | 73.53 | 84.58 | 60.53 |
| MAML based | 64.60 | 52.44 | 33.67 | 27.31 |
| MN based | 78.28 | 61.72 | 68.86 | 48.30 |
| TDNN based | 90.77 | 74.64 | 85.35 | 64.73 |
| Res2Net based | 89.51 | 76.86 | 82.06 | 66.43 |
| PCA based | 91.36 | 74.41 | 86.12 | 62.74 |
| Filter based | 81.36 | 64.00 | 69.68 | 51.90 |

TABLE VII
ACCURACY SCORES OBTAINED BY DIFFERENT METHODS ON THE V1-SET (IN %)

| Methods | 5-way 5-shot | 5-way 1-shot | 10-way 5-shot | 10-way 1-shot |
|---|---|---|---|---|
| Ours | **92.74** | **77.63** | **88.43** | 66.37 |
| MN1D based | 92.27 | 76.68 | 88.24 | **67.01** |
| GC based | 92.02 | 76.13 | 87.69 | 66.93 |
| ETL based | 90.22 | 73.31 | 84.75 | 63.06 |
| PN based | 89.88 | 73.19 | 82.17 | 59.50 |
| MAML based | 65.64 | 53.24 | 42.99 | 32.38 |
| MN based | 77.79 | 59.58 | 68.19 | 45.51 |
| TDNN based | 90.61 | 74.15 | 84.53 | 63.73 |
| Res2Net based | 85.04 | 72.99 | 75.31 | 60.35 |
| PCA based | 87.11 | 69.06 | 79.17 | 57.17 |
| Filter based | 78.33 | 61.68 | 65.96 | 46.18 |

TABLE VIII
ACCURACY SCORES OBTAINED BY DIFFERENT METHODS ON THE L-SET (IN %)

| Methods | 5-way 5-shot | 5-way 1-shot | 10-way 5-shot | 10-way 1-shot |
|---|---|---|---|---|
| Ours | **98.51** | **92.99** | **97.59** | **90.18** |
| MN1D based | 98.39 | 92.96 | 97.48 | 89.62 |
| GC based | 98.29 | 92.73 | 97.46 | 89.52 |
| ETL based | 98.28 | 91.61 | 97.04 | 87.03 |
| PN based | 97.67 | 87.07 | 94.61 | 79.86 |
| MAML based | 76.01 | 66.75 | 65.15 | 48.91 |
| MN based | 95.39 | 86.56 | 92.63 | 79.12 |
| TDNN based | 97.93 | 90.51 | 96.50 | 86.49 |
| Res2Net based | 97.36 | 90.40 | 93.15 | 84.48 |
| PCA based | 96.83 | 87.33 | 94.68 | 80.25 |
| Filter based | 96.04 | 83.84 | 92.31 | 73.59 |

As shown in Table VI, our method obtains accuracy scores of 92.89%, 78.31%, 88.87% and 68.88% on the V2-set when the values of $N$-way $K$-shot are set to 5-way 5-shot, 5-way 1-shot, 10-way 5-shot, and 10-way 1-shot, respectively. These accuracy scores are higher than the counterparts produced by the baseline methods. Hence, our method exceeds baseline methods in accuracy when evaluated on the V2-set. The same conclusions can be made from the results of Table VIII. When different methods are evaluated on the V1-set, as shown in Table VII, the accuracy scores obtained by our method are higher than the counterparts obtained by other methods, except the case of 10-way 1-shot. In summary, our method exceeds all baseline methods on three datasets across different settings of $N$-way $K$-shot, except the case of 10-way 1-shot on the V1-set. The advantage of our method in accuracy over the baseline methods mainly benefits from the designs of feature interaction operation and the RCB in the embedding module.

In addition, the accuracy scores obtained by different methods on the L-set are consistently higher than that on the V2-set and V1-set. The reason is probably that the background noise in the L-set is much lower than that in the V2-set and V1-set. Hence, the variations of time-frequency properties among the speech samples of the same speaker from the L-set are smaller than that among the speech samples of the same speaker from the V2-set and V1-set.

*F. Comparison on Truncated Segments in Accuracy*

In this subsection, we compare the performance of different methods on the truncated testing segments with various lengths. Each truncated segment is generated by randomly splitting each testing speech sample (7 seconds) into speech segments with lengths of 1 second, 3 seconds or 5 seconds. These truncated segments in the V2-set, V1-set, and L-set, are adopted as testing data for evaluating the robustness of different methods to the length of truncated testing segments. In this experiment, we discuss the performance of different methods when the value of $N$-way $K$-shot is set to 5-way 5-shot without loss of generality. Table IX lists accuracy scores obtained by different methods on the testing segments with different durations, in which "whole" represents the length of the entire testing speech sample, namely 7 seconds.

TABLE IX
ACCURACY SCORES OBTAINED BY DIFFERENT METHODS ON TRUNCATED TESTING SEGMENTS (IN %)

| Methods | Length | V2-set | V1-set | L-set |
|---|---|---|---|---|
| Ours | 1 s | 87.14 | 82.00 | 96.69 |
|  | 3 s | 91.99 | 89.80 | 97.97 |
|  | 5 s | 92.80 | 92.51 | 98.27 |
|  | whole | **92.89** | **92.74** | **98.51** |
| MN1D based | 1 s | 80.98 | 77.08 | 94.99 |
|  | 3 s | 89.71 | 88.70 | 97.84 |
|  | 5 s | 91.81 | 91.49 | 97.86 |
|  | whole | 92.27 | 92.27 | 98.39 |
| GC based | 1 s | 81.44 | 76.48 | 95.19 |
|  | 3 s | 89.82 | 88.46 | 97.81 |
|  | 5 s | 91.22 | 91.60 | 98.19 |
|  | whole | 92.33 | 92.02 | 98.29 |
| ETL based | 1 s | 73.66 | 70.32 | 94.40 |
|  | 3 s | 86.47 | 83.52 | 97.97 |
|  | 5 s | 89.15 | 89.25 | 98.01 |
|  | whole | 90.83 | 90.22 | 98.28 |
| PN based | 1 s | 76.01 | 75.61 | 96.07 |
|  | 3 s | 86.92 | 86.27 | 96.67 |
|  | 5 s | 89.87 | 89.33 | 97.63 |
|  | whole | 90.47 | 89.88 | 97.67 |
| MAML based | 1 s | 54.70 | 51.57 | 67.32 |
|  | 3 s | 62.55 | 61.95 | 74.35 |
|  | 5 s | 63.85 | 63.95 | 75.62 |
|  | whole | 64.60 | 65.64 | 76.01 |
| MN based | 1 s | 58.95 | 53.17 | 87.46 |
|  | 3 s | 71.47 | 68.90 | 92.40 |
|  | 5 s | 76.84 | 75.11 | 93.88 |
|  | whole | 78.28 | 77.79 | 95.39 |
| TDNN based | 1 s | 77.01 | 72.34 | 93.52 |
|  | 3 s | 87.57 | 86.41 | 96.99 |
|  | 5 s | 90.61 | 89.60 | 97.63 |
|  | whole | 90.77 | 90.61 | 97.93 |
| Res2Net based | 1 s | 80.70 | 72.53 | 95.09 |
|  | 3 s | 87.02 | 81.82 | 96.39 |
|  | 5 s | 87.82 | 84.07 | 96.89 |
|  | whole | 89.51 | 85.04 | 97.36 |
| PCA based | 1 s | 80.79 | 72.42 | 92.35 |
|  | 3 s | 89.97 | 85.45 | 95.18 |
|  | 5 s | 91.17 | 86.87 | 95.98 |
|  | whole | 91.36 | 87.11 | 96.83 |
| Filter based | 1 s | 76.14 | 69.70 | 92.78 |
|  | 3 s | 80.04 | 77.84 | 95.30 |
|  | 5 s | 80.82 | 78.08 | 95.87 |
|  | whole | 81.36 | 78.33 | 96.04 |

Based on the accuracy scores obtained by different methods in Table IX, the following four observations can be obtained.

First, the accuracy scores obtained by all methods on all testing subsets constantly decrease with the decrease of the lengths of testing speech segments. Furthermore, the decrease of accuracy scores produced by our method is smaller than that obtained by most baseline methods. For example, when the lengths of speech segments in the V2-set decrease from 5 seconds to 1 second, the absolute reduction of the accuracy score achieved by our method is 5.66% (92.80% - 87.14%). This value (5.66%) is smaller than the counterparts produced by all baseline methods, except the filter based method (80.82% - 76.14% = 4.68%).

Second, the shorter the testing speech segment is, the smaller the accuracy scores obtained by different methods are. For example, the accuracy scores obtained by our method decrease from 92.89% to 87.14% when the lengths of testing speech segments in the V2-set decrease from "whole" to 1 second. Similar results are obtained for the baseline methods when they are evaluated on different testing subsets.

Third, our method outperforms all baseline methods in accuracy when they are evaluated on truncated segments with different lengths. For example, our method obtains the highest accuracy score of 87.14% when evaluated on the speech segments with 1 second in the V2-set. However, the maximum of the corresponding accuracy scores obtained by the baseline methods is 81.44%. The same observations can be obtained when our method and baseline methods are evaluated on the speech segments with different lengths in all testing subsets.

Fourth, the shorter the length of testing speech segment is, the larger the accuracy margins between our method and most of the baseline methods are. For instance, the absolute margin of accuracy score between our method and the MAML based method is 28.29% (92.89% - 64.60%), when these two methods are evaluated on the speech segments with length of "whole" in the V2-set. However, the counterpart between these two methods becomes 32.44% (87.14% - 54.70%) when the speech segments with length of 1 second are adopted as testing data.

In conclusion, our proposed method still exceeds the baseline methods when they are assessed on the truncated testing segments with various lengths in terms of accuracy. In addition, our method is robust to the length of the truncated testing segments, since it still produces higher accuracy scores on the truncated testing segments. The reason is that the speaker embedding learned by the proposed embedding module can effectively represent both the global sequential information and the local spatial information. Accordingly, our proposed method generalizes well across truncated testing segments with different lengths instead of overfitting on the segments with single length.

### G. Comparison of Different Methods in Complexity

In this subsection, we measure the memory requirements of different methods using the metric of MS. In addition, computational complexities of different methods are measured by the metric of MACs when the lengths of speech segments are equal to 1 second, 3 seconds and 5 seconds. The values of MS and MACs of different methods are presented in Table X.

TABLE X
MEMORY REQUIREMENTS AND COMPUTATIONAL COMPLEXITIES OF DIFFERENT METHODS.

| Methods | MS (Kilo) | MACs (Million) | | |
|---|---|---|---|---|
| | | 1 s | 3 s | 5 s |
| Ours | **54.14** | **5.54** | **16.63** | **27.71** |
| MN1D based | 59.26 | 6.05 | 18.16 | 30.27 |
| GC based | 79.10 | 8.04 | 24.12 | 40.19 |
| ETL based | 364.75 | 12.39 | 36.94 | 61.48 |
| PN based | 179.71 | 7.42 | 22.31 | 37.20 |
| MAML based | 245.54 | 23.77 | 73.53 | 124.05 |
| MN based | 111.77 | 104.79 | 314.76 | 523.30 |
| TDNN based | 6185.44 | 467.65 | 1399.88 | 2332.11 |
| Res2Net based | 4418.58 | 13.99 | 24.06 | 34.12 |
| PCA based | 164.35 | 5.88 | 17.70 | 29.52 |
| Filter based | 164.35 | 5.88 | 17.70 | 29.52 |

Note: The values of MS and MACs of only the back-end model (namely prototypical network) are included for the PCA based and filter based methods, since there is no front-end model for feature learning in these two methods. Hence, the values of MS and MACs of these two methods are the same.

In terms of memory requirement, the MS of our proposed method is 54.14 kilo which is smaller than that of all baseline methods. In terms of computational complexity, the values of the MACs of our proposed method are 5.54 million, 16.63 million and 27.71 million on speech segments with 1 second, 3 seconds and 5 seconds, respectively. Moreover, the values of the MACs of our method are lower than the counterparts of all baseline methods, when they are evaluated on speech segments with different lengths.

In summary, our proposed method has advantage over all baseline methods in terms of both memory requirement and computational complexity. The advantages of our proposed method in these two aspects over all baseline methods mainly benefit from the designs of feature grouping and the DRC block in the proposed embedding module.

### H. Generalization across Datasets

In all experiments above, the training subset and testing subset are chosen from the same dataset. To evaluate the generalization performance of various methods across datasets, the training subset and the testing subset are from different datasets. That is, when the training subset is from a dataset (e.g., V2-set), the testing subset is from the remaining two datasets (e.g., V1-set and L-set). In this experiment, the value of $N$-way $K$-shot is set to 5-way 5-shot without loss of generality, and each input sample is a whole speech sample.

In the first row of Table XI, the item on the left side (e.g., "V2" in "V2→V1") and the item on the right side (e.g., "V1" in "V2→V1") of the arrow denote the training subset and the testing subset, respectively. The accuracy scores obtained by different methods across datasets are listed in Table XI.

TABLE XI
ACCURACY SCORES OBTAINED BY VARIOUS METHODS ACROSS DATASETS (IN %)

| Methods | V2→V1 | V2→L | V1→V2 | V1→L | L→V2 | L→V1 |
|---|---|---|---|---|---|---|
| Ours | **92.65** | **99.41** | **92.17** | 98.41 | 87.52 | **88.32** |
| MN1D based | 92.15 | 98.51 | 91.80 | **98.51** | **88.01** | 88.26 |
| GC based | 92.34 | 98.58 | 91.74 | 98.32 | 87.49 | 88.07 |
| ETL based | 87.15 | 98.00 | 90.34 | 96.79 | 84.36 | 84.64 |
| PN based | 88.68 | 96.37 | 89.74 | 96.94 | 78.61 | 78.91 |
| MAML based | 63.42 | 70.00 | 66.21 | 73.15 | 63.38 | 63.14 |
| MN based | 75.62 | 88.96 | 79.39 | 90.72 | 72.11 | 69.74 |
| TDNN based | 89.83 | 96.72 | 90.11 | 97.37 | 80.38 | 80.29 |
| Res2Net based | 88.50 | 95.72 | 84.49 | 93.09 | 80.01 | 80.11 |
| PCA based | 89.64 | 96.14 | 88.68 | 95.54 | 79.67 | 77.48 |
| Filter based | 80.23 | 86.44 | 78.33 | 86.50 | 77.81 | 78.00 |

V2: V2-set; V1: V1-set; L: L-set.

Our proposed method obtains accuracy scores of 92.65%, 99.41%, 92.17%, 98.41%, 87.52%, and 88.32% when datasets are V2→V1, V2→L, V1→V2, V1→L, L→V2, and L→V1, respectively. These accuracy scores are higher than the counterparts obtained by baseline methods, except the cases of V1→L and L→V2. Hence, our proposed method still performs well when the training and testing subsets are from different datasets.

As given in the second row and second column of Tables VI, VII, and VIII, our proposed method obtains accuracy scores of 92.89%, 92.74%, and 98.51% when datasets are V2→V2, V1→V1, and L→L (training and testing subsets from the same datasets), respectively. The accuracy score of 92.89% (V2→V2) is higher than the accuracy score of 92.65% (V2→V1) but is lower than the accuracy score of 99.41% (V2→L). Similarly, the accuracy score of 92.74% (V1→V1) is higher than the accuracy scores of 92.17% (V1→V2), but is lower than the accuracy score of 98.41% (V1→L). However, the accuracy score of 98.51% (L→L) is higher than the accuracy scores of 87.52% (L→V2) and 88.32% (L→V1). In short, our proposed method obtains better performance when the training subset and the testing subset (except the L-set) are from the same datasets. In addition, when the testing subset is the L-set, even if the training subset is different from the testing subset, our proposed method obtains higher accuracy scores. The reason is probably that the speech samples in the L-set are clean (without evident background noise). Accordingly, the distribution of time-frequency properties of speech samples in the L-set is relatively simpler and may overlap with that of speech samples in the V1-set and V2-set.

In summary, our proposed method generalizes well across datasets instead of overfitting on a single dataset.

*I. Comparison of Different Methods for Speaker Verification*

In this subsection, we conduct one extended experiment for comparing different methods on three datasets for speaker verification. In this experiment, the value of *N*-way *K*-shot is set to 5-way 5-shot, and each feature is learned from a whole speech sample. A typical metric of equal error rate (EER) is adopted to measure the performance of all methods for speaker verification. The EER is defined as the rate at which acceptance error equals to rejection error. The lower the EER score is, the better the performance of the methods is.

Under the same conditions, the EER scores that are produced by different methods on the datasets of the V2-set, V1-set and L-set are given in Table XII. Our method obtains EER scores of 12.98%, 15.32% and 8.32% on the V2-set, V1-set and L-set, respectively. These EER scores are lower than the counterparts obtained by all baseline methods. In other words, our proposed method outperforms all baseline methods in terms of EER when they are evaluated on the datasets of the V2-set, V1-set and L-set. Therefore, the proposed embedding module for learning speaker embedding is also effective in the task of speaker verification.

TABLE XII
EER SCORES OBTAINED BY DIFFERENT METHODS ON THREE DATASETS (IN %)

| Methods | V2-set | V1-set | L-set |
|---|---|---|---|
| **Ours** | **12.98** | **15.32** | **8.32** |
| MN1D based | 13.98 | 16.98 | 10.99 |
| GC based | 13.65 | 15.98 | 11.99 |
| ETL based | 18.31 | 21.31 | 12.65 |
| PN based | 21.97 | 27.31 | 20.98 |
| MAML based | 44.95 | 46.29 | 39.29 |
| MN based | 25.97 | 28.31 | 21.98 |
| TDNN based | 15.98 | 17.98 | 16.32 |
| Res2Net based | 14.65 | 20.64 | 10.66 |
| PCA based | 16.98 | 19.31 | 18.65 |
| Filter based | 24.31 | 25.97 | 20.97 |

## VI. CONCLUSIONS

In this work, we have investigated a newly emerging problem of lightweight FSSI. Moreover, we have tackled this problem by designing a lightweight prototypical network with the operations of both feature grouping and feature interaction. Based on the description of our proposed method and the discussions of experimental results, we can draw the following three conclusions.

First, our propose method basically outperforms baseline methods in accuracy under many experimental conditions, such as variable numbers of *N*-way *K*-shot, various lengths of speech segments, different datasets. Hence, our proposed method is a state-of-the-art method for solving the problem of lightweight FSSI.

Second, the proposed method has advantage over all baseline methods in terms of memory requirement and computational complexity. In addition, the minimum margins of the values of both MS and MACs between the proposed method and the baseline methods are 5.12 (59.26 - 54.14) kilo and 0.34 (5.88 - 5.54) million, respectively.

Third, we design a computationally-efficient embedding module which is used to learn speaker embedding with strong representational ability. The proposed embedding module acquires both global sequential information and local spatial information, and carries out the operations of both feature grouping and feature interaction. The acquisition of the two kinds of information above and the operation of feature interaction have enhanced the representation ability of the learned speaker embedding. The designs of the DRC block and the operation of feature grouping have been beneficial for reducing the computational complexity and model size of the proposed prototypical network. The extended experiments have further verified our proposed method's robustness to the length of truncated testing segments, the generalization across datasets, and the effectiveness for speaker verification.

Although our proposed method has obtained satisfactory results for lightweight FFSI, there are still some areas for improvement in our proposed method. For example, the BLSTM block in the RCB is not implemented in parallel, which obviously increases the computational complexity of the proposed embedding module. In addition, we do not consider the attention mechanism in the proposed embedding module, which affects the representational ability of the learned speaker embedding as well. In the next work, we will optimize the architecture of the proposed network to further reduce the computational complexity and enhance the representational ability of speaker embedding by proposing extra effective strategies or modules. Specifically, we will consider designing an attention block which can be implemented in parallel for replacing the BLSTM block. Other techniques, such as network quantization, linear transformation with higher computational efficiency, will be considered as well. Accordingly, the proposed network will become lighter and perform better for deploying the proposed method on the intelligent terminals with limited resources.